\documentclass[aip,apl,amsmath,amssymb,reprint]{revtex4-1}
\usepackage[T1]{fontenc} 
\usepackage{graphicx}
\usepackage{dcolumn}
\usepackage{bm}
\usepackage{comment}
\usepackage{soul}
\usepackage[dvipsnames]{xcolor}
\usepackage{color}
\usepackage{tabularx}
\newcolumntype{Y}{>{\centering\arraybackslash}X}
\usepackage{booktabs}

\usepackage[mathlines]{lineno}

\begin{document}


\title{Extensive study of electron acceleration by relativistic surface plasmons}%
\author{G. Cantono}
\email{giada_cantono@hotmail.it}
\affiliation{LIDYL, CEA, CNRS, Universit\'e Paris-Saclay, CEA Saclay, 91191 Gif-sur-Yvette, France}
\affiliation{Universit{\'e} Paris Sud, Paris, 91400 Orsay, France}
\affiliation{National Institute of Optics, National Research Council (CNR/INO) A. Gozzini unit, 56124 Pisa, Italy}
\affiliation{Enrico Fermi department of Physics, University of Pisa, 56127 Pisa, Italy} 
\author{A. Sgattoni}
\affiliation{LULI-UPMC: Sorbonne Universit\'es, CNRS,  \'Ecole Polytechnique, CEA, 75005 Paris, France}
\affiliation{LESIA, Observatoire de Paris, CNRS, UPMC: Sorbonne Universites, 92195 Meudon, France}
\affiliation{National Institute of Optics, National Research Council (CNR/INO) A. Gozzini unit, 56124 Pisa, Italy}
\author{L. Fedeli}
\affiliation{Department of Energy, Politecnico di Milano, 20133 Milano, Italy}
\author{D. Garzella}
\affiliation{LIDYL, CEA, CNRS, Universit\'e Paris-Saclay, CEA Saclay, 91191 Gif-sur-Yvette, France}
\author{F. R\'eau}
\affiliation{LIDYL, CEA, CNRS, Universit\'e Paris-Saclay, CEA Saclay, 91191 Gif-sur-Yvette, France}
\author{C. Riconda}
\affiliation{LULI-UPMC: Sorbonne Universit\'es, CNRS,  \'Ecole Polytechnique, CEA, 75005 Paris, France}
\author{A. Macchi}
\affiliation{National Institute of Optics, National Research Council (CNR/INO) A. Gozzini unit, 56124 Pisa, Italy}
\affiliation{Enrico Fermi department of Physics, University of Pisa, 56127 Pisa, Italy}
\author{T. Ceccotti}
\affiliation{LIDYL, CEA, CNRS, Universit\'e Paris-Saclay, CEA Saclay, 91191 Gif-sur-Yvette, France}

\date{29 November 2017}

\begin{abstract}
The excitation of surface plasmons with ultra-intense ($I\sim 5\times 10^{19}$ W/cm$^2$), high contrast ($\sim 10^{12}$) laser pulses on periodically-modulated solid targets has been recently demonstrated to produce collimated bunches of energetic electrons along the target surface [Fedeli et al., Phys. Rev. Lett. 116, 5001 (2016)]. 
Here we report an extensive experimental and numerical study aimed to a complete characterization  of the acceleration mechanism, demonstrating its robustness and promising characteristics for an electron source. By comparing different grating structures, we identify the relevant parameters to optimize the acceleration and obtain bunches of $\sim 650$ pC of charge at several MeV of energy with blazed gratings.

\end{abstract}

\maketitle
\section*{Introduction}
Plasmonics in the relativistic regime is a daring but largely unexplored domain. Exploiting the unique properties of light confinement\cite{Barnes2003, Schuller2010} and field concentration \cite{Gramotnev2014} achieved with the excitation of surface plasmons (SPs) on metallic nano-structures could open new schemes of laser-plasma interaction at high field intensity and the possibility to improve laser-based radiation sources, which would notably profit from enhancing the laser-target coupling. 

Indeed, SPs have been studied for the last few decades to efficiently increase the absorption of the laser energy by an overdense plasma. Exciting SPs at high laser intensities \cite{Macchi2017, Fedeli2016b} can generate very strong fields close to the surface and, in turn, produce a numerous population of highly-energetic electrons and enhance the emission of both protons \cite{Bigongiari2013, Ceccotti2013} and XUV harmonics from the target \cite{Fedeli2017}. 

Although promising, there is no decisive theory of SP excitation by ultra-intense laser pulses at the interface of a solid plasma, whose strongly non-linear response cannot be described by a univocal dielectric function because of the relativistic effects\cite{Weng2012}. 
Nevertheless, numerical simulations \cite{Raynaud2007, Bigongiari2011} soon encouraged the possibility to excite SPs on grating targets irradiated at relativistic intensities at the proper resonant angle predicted by the linear, non-relativistic theory \cite{Maier2004}. Yet, the first experiments on this topic\cite{Kahaly2008, Hu2010a,Bagchi2012} were limited to intensities far below $10^{18}$ W/cm$^2$, because the poor temporal contrast inherent to powerful CPA laser systems did not ensure the survival of the grating surface, irradiated by the pulse pedestal before the arrival on target of the main intensity peak. 
The successful development of pulse cleaning techniques, such as the plasma mirror \cite{Thaury2007}, recently enabled the experimental study of plasmonic effects at relativistic laser intensities.

In this context, recent experiments reported not only SP-enhanced proton acceleration \cite{Ceccotti2013}, but the remarkable acceleration of electron bunches along the target surface directly driven by the SP electric field \cite{Fedeli2016, Sgattoni2015}. This process, which had been only partly investigated by previous numerical simulations\cite{Raynaud2007, Bigongiari2011}, has been thoroughly addressed in a series of experiments performed at CEA Saclay \cite{Cantono2017}. 

In this paper we report the exhaustive description of both experimental and numerical results, aiming to demonstrate the peculiar features of the electron emission and its dependence on some of the target and laser parameters. These results emphasize the robustness of the acceleration mechanism and encourage the development of a compact electron source at few MeV of energy, with potential applications for ultra-fast electron diffraction \cite{Zhu2015, Tokita2009}, photo-neutron generation \cite{Pomerantz2014} or enhanced emission of THz radiation \cite{Fedeli2018}. 

\section{Surface plasmons for electron acceleration} \label{sec-theory}
\begin{figure}[tbp]
\includegraphics[width=1.0\columnwidth]{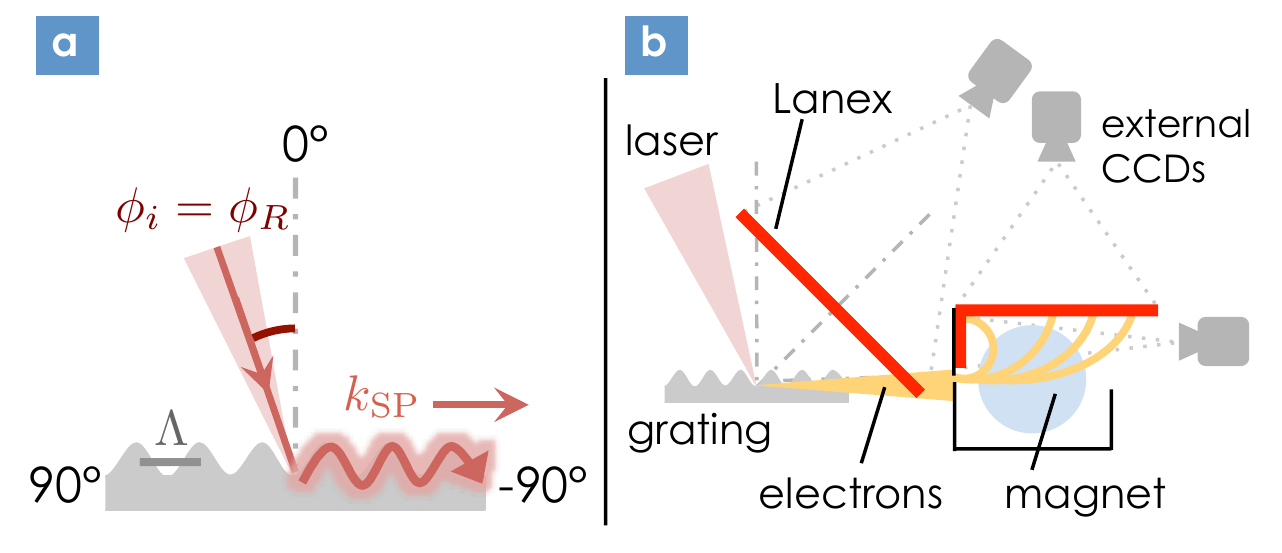}
\caption{(a) Scheme of the SP excitation on a grating, with the sign convention used in the text. The incidence angle $\phi_i$ corresponds to the resonant angle fixed by Eq.~\ref{res-cond}. \\ (b) Layout of the experiments performed at CEA Saclay. In the spectrometer, the electron trajectory describing half a circle corresponds to $1.4$ MeV of energy. The Lanex inside the yoke, reserved for electrons below this value, never produced a detectable signal.}
\label{fig1-scheme} 
\end{figure}

SPs are normal modes of the electronic oscillations at a sharp metal-dielectric interface. 
They can be excited by an external laser pulse on a periodically-modulated target that achieves phase-matching with the incoming electromagnetic wave\cite{Maier2004, Sgattoni2015}. In the relativistic regime, the solid target is ionized within a laser cycle, allowing the rest of the laser pulse to interact with an overdense plasma. A short pulse duration ($\sim 10$s of fs) and high contrast are required to avoid both the early smoothing and the following hydrodynamic expansion that otherwise would destroy the modulated surface of the target. If this one consists in a diffraction grating, resonance occurs when the laser pulse irradiates the grating at a specific incidence angle $\phi_R$, related to the grating period $\Lambda$ by the condition:
\begin{equation} \label{res-cond} \sin{(\phi_R)}= \pm \sqrt{\frac{1-\omega_p^2/\omega^2}{2-\omega_p^2/\omega^2}} + n\frac{\lambda}{\Lambda}. \end{equation}
In this expression, $\omega_p$ is the plasma frequency, $\omega$ and $\lambda$ are the frequency and wavelength of the laser pulse, and $n$ is an integer ($0$, $\pm 1$, ...).
The first term on the right-hand side represents the SP dispersion relation $ck_\text{SP}/\omega$ for a cold, collisionless plasma \cite{Kaw1970} derived from the linear theory. However, Eq.~\ref{res-cond} can be reduced to $\sin{(\phi_R)}= \pm 1 + n\lambda/\Lambda$ in the limit of solid targets, since $\omega_p \gg \omega$ holds due to the high electron density. Measuring the angles from the target normal as indicated in Fig.~\ref{fig1-scheme}a, this expression reminds of the well-known grating equation, relating the incidence at $\phi_R$ of a monochromatic beam to the propagation of the $n$-th diffraction order along the grating surface. Indeed, SPs are particular solutions of the EM field diffracted by a grating, characterized by the field confinement in the direction perpendicular to the interface \cite{McDonald2000}.

Electrons can be accelerated along the grating by the component of the SP electric field that is parallel to the surface, $E_\parallel$. Efficient acceleration occurs on the vacuum side of plasma-vacuum interface, since the SP field is rapidly evanescent inside the plasma: electrons are pulled into the vacuum region by the transverse component of the electric field, $E_\perp$, then self-injected in the SP field thanks to the  $\mathbf{J}\times \mathbf{B}$ force directed along the surface \cite{Riconda2015, Fedeli2016}. 
An example of electron trajectory super-imposed on the temporal evolution of electric field during the SP excitation is given in the supplemental material. 

The phase velocity of the SP $v_\varphi=\omega/k_\text{SP}$ does not depend on the SP intensity, and for $\omega_p \gg \omega$, it is $v_\varphi \sim c$. Therefore, relativistic laser intensities are required to accelerate the electrons close to $v_\varphi$. Then, depending on the initial conditions\cite{Fedeli2016, Riconda2015}, 
an electron can attain a kinetic energy $W \simeq m_ec^2 \gamma_\varphi a_\text{SP} \gg m_ec^2$, where $\gamma_\varphi = (\omega_p^2/\omega^2-1)^{1/2}$ is the relativistic factor associated to the phase velocity of the SP and $a_\text{SP}=eE_\perp/(m_e\omega c)$ is the normalized electron momentum in the transverse electric field of the SP (knowing that $E_\perp = \gamma_\varphi E_\parallel$). Since $v_\varphi \sim c$ and the evanescence length on the vacuum side $L_{e,v}=(k_\text{SP}^2-\omega^2/c^2)^{-1/2}$ exceeds the laser wavelength $\lambda$, the electron can remain in phase with the SP for a long time, despite being steered away from the surface by $E_\perp$. Consequently, acceleration lengths $L_{acc}=W/(eE_\parallel)=\gamma_\varphi^2/k$ can be achieved \cite{Cantono2017}.

Notice that because of self-injection, a large amount of charge can be synchronized to the  accelerating field of the SP and reach high energies. In this way, this acceleration mechanism quite differs from the dielectric laser acceleration\cite{Peralta2013} or from the inverse Smith-Purcell effect\cite{Mizuno1975}, where an external electron beam injected at grazing incidence on a periodic structure can be accelerated by the field induced on the micro-structure by a low-intensity, ps laser pulse. 
These processes do not involve the excitation of a SP and acceleration is achieved only if the  electron beam is carefully synchronized to laser field. In the inverse Smith-Purcell effect, such synchronous condition should not be mistaken with the resonant condition Eq. \ref{res-cond}, which derives from the phase-matching of the SP and the laser pulse and does not relate to the velocity of the accelerated electrons.
Moreover, the inverse Smith-Purcell acceleration requires the laser beam to hit the grating at skew incidence\cite{Palmer1980}, whereas tilting the grating lines or varying the incidence angle spoils the electron acceleration in our case.

For a comparison with the experimental measurements reported in the following, we can derive $E_\perp$ by assuming that the laser energy deposited in the focal spot is entirely yielded to the SP. Despite leading to an overestimation of  $a_\text{SP}$, this choice is supported by previous measurements of the target absorption, where values near $100$\% were reported \cite{Kahaly2008, Ceccotti2013}. In this way, $a_\text{SP}$  becomes $\sim 3$ for the peak intensity of the laser system UHI-100 described in the next section. Assuming a solid target density of $n_e \sim 400 n_c$, the theoretical model predicts a $\gamma_\varphi$ factor of $\sim 20$, hence a maximum kinetic energy of $W \sim 30$ MeV. The emission angle, measured on the incidence plane from the grating normal, is expressed\cite{Fedeli2016} by $\tan(\phi)=\gamma_\varphi \beta_\varphi$ , resulting here in $\sim 87^\circ$. The threshold value $a_\text{SP}\sim1$, which was also recovered from 2D simulations\cite{Fedeli2016}, still leads to a maximum electron energy $W \sim 10$ MeV $\gg m_ec^2$.

\section{Experimental results} \label{sec-experiments}
Experiments were carried out at the Saclay Laser-matter Interaction Center Facility (SLIC) of CEA Saclay (Gif sur Yvette, France). The UHI-100 Ti:Sa laser system delivers $25$ fs pulses with $\sim 2.5$ J of energy before compression. The spectrum is centered at $\lambda \simeq 800$ nm with $80$ nm of FWHM bandwidth. 
A double plasma mirror increases the temporal contrast to $10^{12}$ and $10^{10}$ within, respectively, $\sim 20$ and $5$ ps before the pulse peak\cite{Levy2007}. Wavefront correction is performed by a deformable mirror, allowing the P-polarized beam to be focused on target at $\sim 4.6$ $\mu$m FWHM with a $f/3.75$ off-axis parabola. 
The energy on target is estimated to be $\sim 700$ mJ, corresponding to a peak intensity ranging from $3.4$ to $1.7 \times 10^{19}$ W/cm$^2$ depending on the incidence angle $\phi_i$, which was varied between $10^\circ$ and $60^\circ$ by properly rotating the target along its vertical axis. 

We used different types of gratings in order to explore the SP-driven electron acceleration under various conditions. Thin gratings, with a sinusoidal profile, were produced by heat-embossing $13$ $\mu$m thick Mylar$^\text{TM}$ foils with a metallic master (HoloPlus, CZ). 
The grating periods $\Lambda$ were $1.35\lambda - 2\lambda-3.41\lambda$ for a resonant angle of, respectively, $15^\circ-30^\circ-45^\circ$ according to Eq.~\ref{res-cond}; these targets are hence referred to as G$15,~$G$30$ and G$45$. The groove depth $d$ was, respectively, $170-290-390$ nm. Depending of the grating type and incidence angle, the number of grating periods irradiated at resonance within the focal spot (at $1/e^2$) was respectively $7.5-5.5-4$.
For the resonance at $30^\circ$ of incidence (\textit{i.e.}~$\Lambda=2\lambda$), we also employed blazed gratings (Edmund Optics). They were produced by depositing float glass on a sawtooth master coated with a $\sim 1$ $\mu$m thick Aluminum layer. The whole thickness of the target was $9.5$ mm and five different blaze angles were tested: $4^\circ-6^\circ-13^\circ-22^\circ-28^\circ$ with, consequently, a groove depth $d$ of $120-180-365-580$ and $700$ nm. These targets will be indicated with the acronym BG, followed by the blaze angle (\textit{e.g.}~BG$13$). In order to clarify the role of the Aluminum coating on the target efficiency, also a thin sinusoidal grating with a period of $2\lambda$ was produced on a $12$ $\mu$m thick Aluminized Mylar foil (\textit{i.e.}~G$30_\text{Alu}$). Finally, we irradiated flat Mylar foils of $13$ $\mu$m thickness for comparison.

The experimental arrangement is illustrated in Fig.~\ref{fig1-scheme}b. 
Electron diagnostics consisted of a scintillating Lanex screen and an electron spectrometer, and were designed to record both the spatial and energetic distribution of the electrons emitted in the half-space in front of the target. The Lanex screen ($15 \times 7$ cm$^2$) was tilted by $45^\circ$ to intercept the electron emission from the tangent to the normal of the target; the distance between the screen and the tangent was $8$ cm. A $200$ $\mu$m thick Aluminum slab was placed in front of the screen to filter out  X-rays and electrons below $\sim 150$ keV. 
In the following, the angular directions along the Lanex screen will be indicated as $\phi$ on the incidence plane and $\theta$ in the vertical direction (the azimuthal and polar angle, respectively). 
The electron spectrometer was aligned $3$ cm behind a $2$ mm diameter hole drilled in the tangent direction of the Lanex screen. It was formed by a pair of round magnets ($0.9$ T of magnetic field) and a collimating slit of $1.5$ mm, which determined a spectral resolution of $\sim 500$ keV for electron energies of $\sim 10$ MeV. The trajectories of electrons above $1.4$ MeV of energy were bent by $90^\circ$ before reaching another Lanex screen protected by $100$ $\mu$m of Aluminum. Electrons with lower energy were curved back towards a third small Lanex placed inside the yoke, right next to the entrance slit. Throughout the experimental campaign, this third screen never produced a detectable signal, suggesting a negligible amount of electrons below $1.4$ MeV. This allowed us to relate the intensity of the signal emitted by the tilted Lanex to the amount of charge reaching its surface, since the energy deposited by electrons in the active layer of the scintillator is independent of their initial energy above $1.5$ MeV\cite{Glinec2006}. To this end, we calibrated both the Lanex and its optical system with a stable electron source provided by the laser-triggered radio-frequency electron accelerator ELYSE (Orsay, France). 

\begin{figure}[bp]
\centering \bigskip
\includegraphics[width=1\columnwidth]{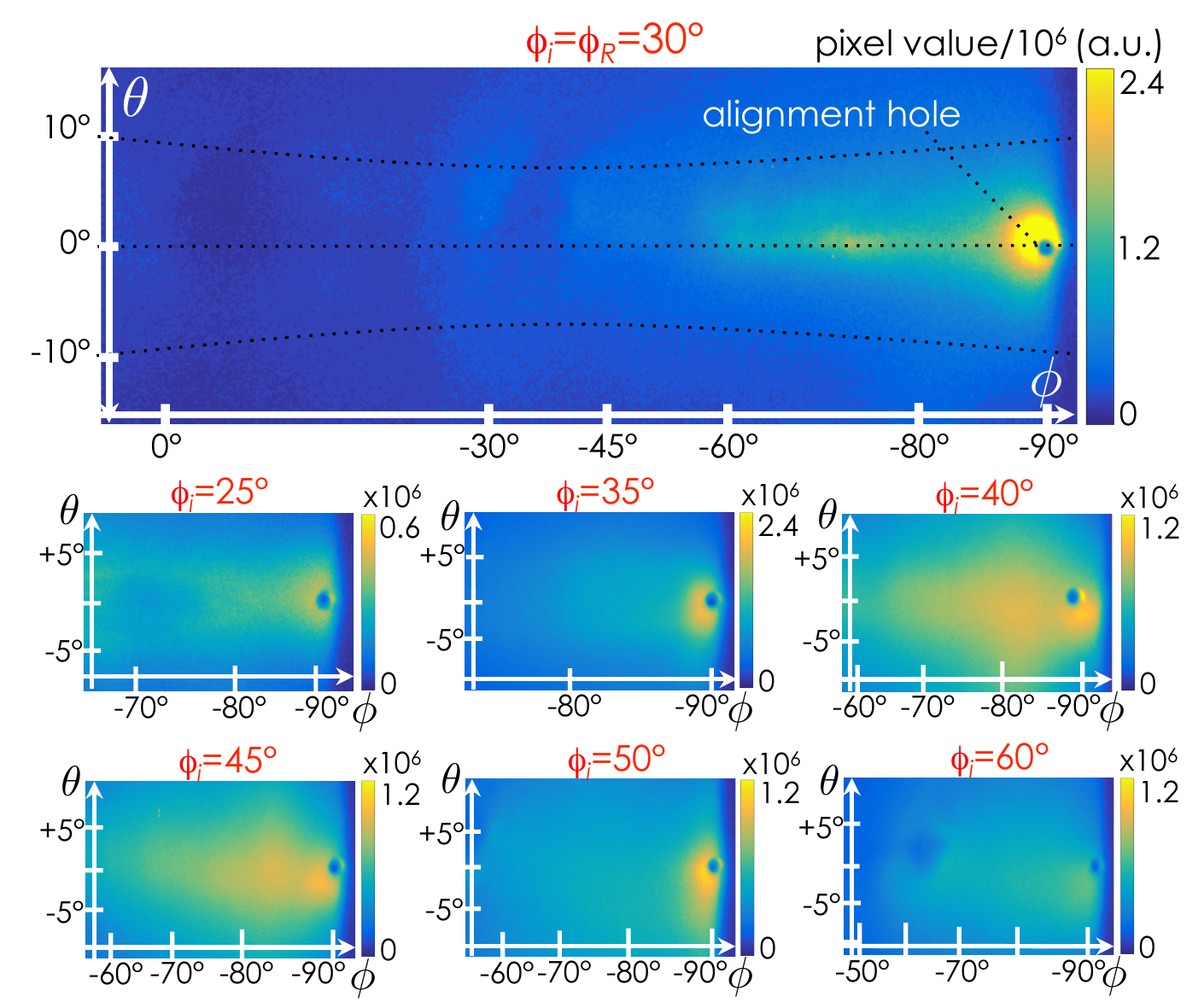}
\caption{\looseness=-1 Electron spatial distribution from a G$30$ irradiated at different incidence angles. The largest panel shows the emission at the resonant angle $\phi_R=30^\circ$, over the entire range of $\phi$ and $\theta$. The angular range is reduced to respectively $\sim 30^\circ$ and $\sim 15^\circ$ in the other images. Color bars indicate the signal intensity. }
\label{fig2-lanex}
\end{figure}
Finally, all diagnostics were mounted on a powered platform which rotated around the chamber center; this allowed us to keep the alignment with the target whenever varying the incidence angle. The signal from the Lanex and electron spectrometer was imaged by $12$-bit CCD cameras equipped with $546$ nm band-pass filters and fixed outside of the interaction chamber. This implied a non-negligible dependence of the measured signal on the distance and the angle between the CCDs and the detectors. Therefore, a correction factor was calculated starting from reference pictures of every position of the rotating platform, to allow for a legitimate comparison between the data acquired on the same points of the Lanex when diagnostics were set at different positions.

\subsection{Electron acceleration for different grating periods} \label{sec-experiments-1}
\looseness=-1 All the gratings irradiated at the expected angle for SP excitation produce an intense, low-divergence electron bunch in the tangent direction, with energies up to $\sim 20$ MeV. Combined with the diameter of the focal spot, these energies suggest that accelerating gradients of $\sim$ TV/m are achieved during the interaction.

Fig.~\ref{fig2-lanex} presents the electron distribution recorded by the Lanex screen from a G$30$ irradiated at various incidence angles. 
The strongest and narrowest emission is found at $\phi_i=\phi_R=30^\circ$; it extends over $\sim 10^\circ$ from the tangent along $\phi$, and over less than $\sim 5^\circ$ along $\theta$. 
Also, two round regions with a weaker signal are observed in the directions corresponding to the specular reflection of the laser pulse ($\phi=-30^\circ$) and to the first diffraction order of this grating ($\phi=0^\circ$, \textit{i.e.}~the target normal)\cite{Fedeli2016}. Similar holes in the specular direction have been reported in other measurements of laser-driven electrons from solid targets\cite{Mordovanakis2009, Thevenet2016}, and attributed to the isotropic scattering exerted by the ponderomotive force of the laser pulse. According to this, the hole at the first diffraction order is a convincing evidence of the grating survival to the pulse pedestal, thanks to the high contrast achieved on UHI-100 \cite{Ceccotti2013,Fedeli2016}.
It is worth mentioning that the flat foil irradiated at the same incidence angle resulted in a $\sim 20$ times weaker signal, with electrons mainly distributed around the specular direction\cite{Fedeli2016}.  

For all the other incidence angles, the electron signal is weaker and spread on a larger area. Notice from Fig.~\ref{fig1-scheme}b that large incidence angles require the platform with the diagnostics to rotate farther away from the CCDs; as a consequence, the same portion of the Lanex screen subtends a wider angular range, especially over $\phi$, and this partially explains why the signal intensity increases around the tangent at large incidence angles. In principle, this effect could also result from a more efficient vacuum heating absorption\cite{Macchi2017}: however, this is in contradiction with the fact that both the measured energetic spectra (compare Fig.~\ref{fig4-spectra}) and the numerical simulations (presented in section \ref{sec-simulations}) indicate that  only low energy electrons (below $\sim 5$ MeV) are emitted in these cases. 

\begin{center}
\begin{figure}[bp]
\centering
\includegraphics[width=1.0\columnwidth]{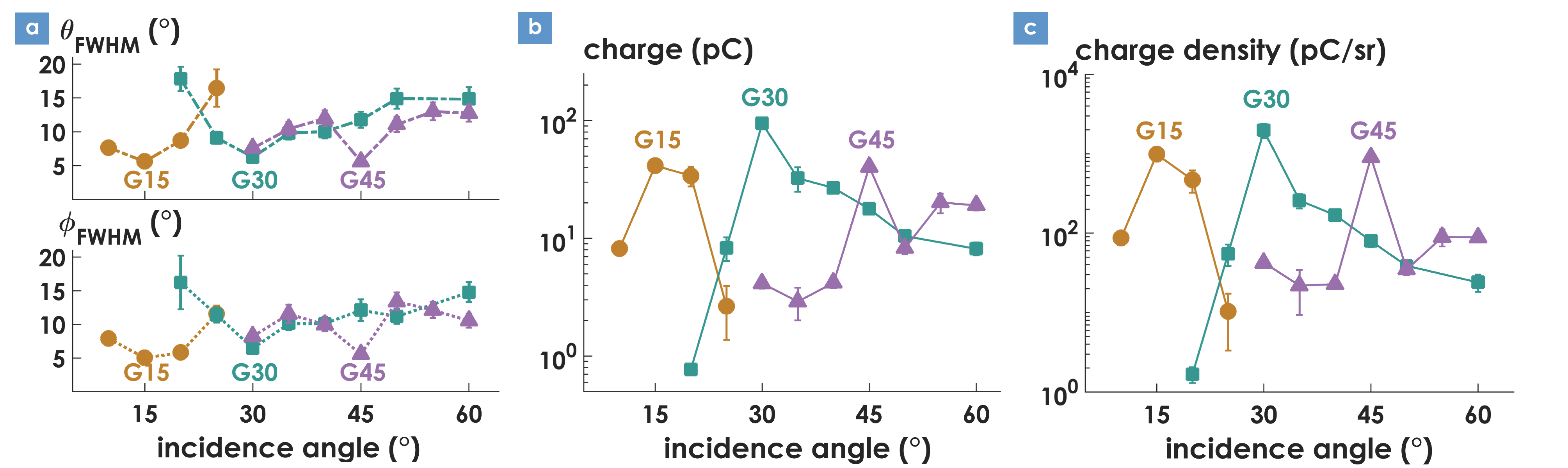}
\caption{Properties of the electron bunch measured along the surface of all thin gratings for different incidence angles: (a) divergence along $\theta$ and $\phi$; (b) charge; (c) charge density. The most intense and collimated bunch is observed at the resonant angle for SP excitation.}  
\label{fig3-spatial-properties}
\end{figure}
\end{center}

\begin{figure}[htbp]
\centering
\includegraphics[width=1.0\columnwidth]{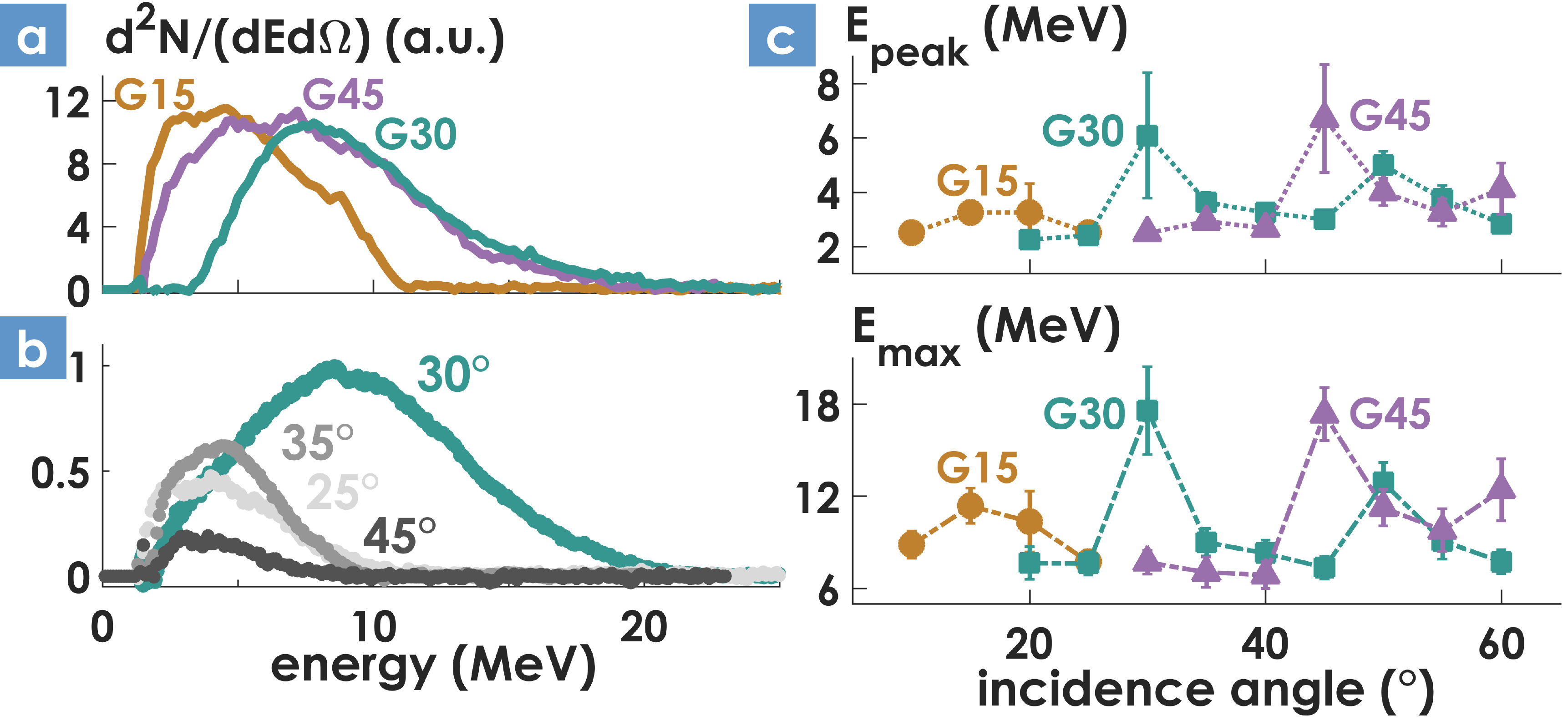}
\caption{Energetic distributions of the electrons accelerated along the surface of thin gratings: (a), energetic spectra collected at the resonant angles; (b), energetic spectra from a G$30$ irradiated at various incidence angles (the spectra are normalized to the peak value of the signal at $30^\circ$); (c), peak and maximum energy for all thin gratings as a function of the incidence angle, emphasizing the effect of the SP excitation at the resonant angles.}
\label{fig4-spectra} 
\end{figure}

Fig.~\ref{fig3-spatial-properties} describes the spatial extent and the charge of the electron bunch inferred from the images of the Lanex screen. The emission along the tangent is analyzed for all thin gratings as a function of the incidence angle. Each point represents the average of all shots acquired in the same configuration, with error bars given by the standard error (\textit{i.e.}~the standard deviation normalized by the square root of the number of shots). When necessary, the standard error is replaced by the systematical error performed during the analysis ($\sim 10$\% of the average); still, in some cases the error bars are hidden by the size of the points shown in the plot.

The angular widths along $\phi$ and $\theta$ are measured on the two orthogonal profiles of the bunch which exhibit the maximum signal. The graph in Fig.~\ref{fig3-spatial-properties}a clearly shows how the electron emission is the less divergent at resonance, with similar FWHM ($\simeq 5^\circ$) for all gratings; the size significantly increases even within $\pm 5^\circ$ of the resonant angle. 
The charge values reported in Fig.~\ref{fig3-spatial-properties}b are estimated inside the area identified for each incidence angle by the FWHM along $\phi$ and $\theta$. The role of the SP is remarkable even in a logarithmic scale, as shown in the graph. Gratings at resonance emit up to $100$ pC of charge, at least $3$ times more than at other incidence angles. Finally, Fig.~\ref{fig3-spatial-properties}c shows the charge density resulting from the combination of both the size and charge of the electron bunch.
\onecolumngrid
\twocolumngrid

\looseness=-1 Fig.~\ref{fig4-spectra}a compares the energetic spectra obtained from all thin gratings at resonance: no electrons above the noise level are detected below $\sim 2$ MeV and the largest population is centered around a peak energy $E_\text{peak}$; the maximum energy, $E_\text{max}$, is measured where the signal $\text{d}N^2/(\text{d}\Omega \text{d}E)$ is equal to $10$\% of its value at the peak. Non-Maxwellian distributions are also found when the gratings are irradiated at non-resonant angles (Fig.~\ref{fig4-spectra}b for a G$30$), yet they show fewer electrons and lower energies. The energetic dispersion $\Delta E/E_\text{peak} \simeq 1.1$ is constant within $5^\circ$ around the resonant angle for all gratings (where $\Delta E$ is the FWHM around $E_\text{peak}$).

The resonant effect is once again visible  in Fig.~\ref{fig4-spectra}c, where $E_\text{peak}$ and $E_\text{max}$ are plotted as a function of the incidence angle. Both the G$30$ and  the G$45$ exhibit similar values, with the maximum energy at resonance around $18$ MeV. The measurements on the G$15$ at resonance, instead, are supposed to have suffered from an accidental misplacement of the electron spectrometer. In fact, the reference pictures of the diagnostics show that the shots at $15^\circ$ of incidence were acquired when the rotating platform was not properly aligned to the target surface. Further support of this hypothesis is given by the simulations, where the maximum electron energy does not appear to depend on the grating type.

For comparison, flat foils were irradiated at the incidence angles corresponding to the grating resonances. In this case, there is no electron acceleration in the tangent direction and electron bunches of variable size are observed in random positions around the specular reflection of the laser beam. For these bunches, Table \ref{tab1-flat-vs-grating} presents the charge values inferred from the images of the Lanex screen, together with the fluctuations of their position in both $\theta$ and $\phi$ directions. These results are compared to the gratings at resonance, which visibly produce $\sim 10$ times more charge and far more directional electron beams. 

Because of the poor reproducibility of the electron emission from flat foils, we did not measure the energetic spectra in the specular direction. Experiments performed with other configurations, involving either very large incidence angles \cite{Li2006} or requiring a pre-formed density gradient at the foil surface \cite{Thevenet2016}, have  reported the acceleration of electrons in the tangent or specular direction from $2$ to $15$ MeV.
\begin{table}[htbp]
\begin{tabularx}{\columnwidth}{YYYYYYY}
$\phi_i(^\circ)$	& \multicolumn{2}{c}{charge (pC)} & \multicolumn{2}{c}{$\Delta \theta (^\circ)$} & \multicolumn{2}{c}{$\Delta \phi(^\circ)$} \\
\midrule
			&F	& 	 G  		&	 F 	& 	 G 	& 	F 	& 	G	\\
$15$		& $3 \pm 1$	& $41 \pm	4$	& $2.1$	& $0.7$	& $2.2$	& $0.2$ \\
$30$	& $4 \pm 1$	& $\mathbf{95} \pm 5$	& $1.8$	& $0.2$	& $2.8$	& $0.05$ \\
$45$ 	& $5 \pm 1$	& $40 \pm 4$	& $2.3$	& $0.3$	& $2.1$	& $0.1$ \\
\end{tabularx}
\caption{Charge and position of the electron bunch emitted from a flat foil (F) around the specular and from a grating (G) along the tangent. The incidence angles correspond to the resonant angles for the gratings. The fluctuations $\Delta \phi$ and $\Delta \theta$ are the standard error of the bunch positions $(\phi,\theta)$ on the data set.}
\label{tab1-flat-vs-grating}
\end{table}

\subsection{Blazed and Aluminized gratings} \label{sec-experiments-2}
Commonly used in low field Plasmonics, blazed gratings (BGs) are designed to maximize the laser energy concentrated into a specific diffraction order (usually $n=1$) for a specific wavelength (known as blaze wavelength) \cite{Palmer2014}.
Ideally, choosing the blaze angle so the that the maximum energy is diffracted along the grating surface should maximize the coupling between the laser pulse and the SP. 
Indeed, with the most suitable BG we found that the charge in the electron bunch increases by $\sim 6$ times with respect to the sinusoidal Mylar gratings. 

\begin{figure}[b]
\includegraphics[width=1.0\columnwidth]{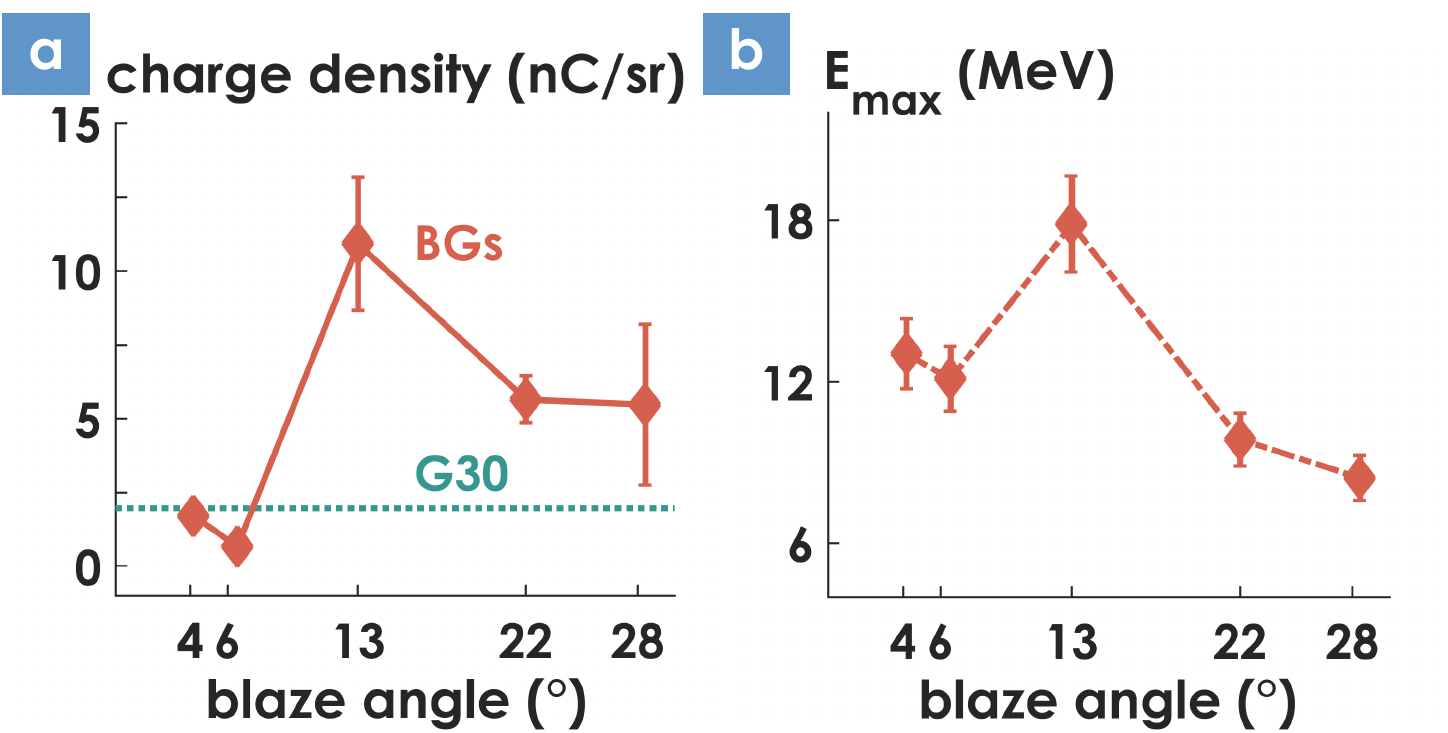}
\caption{Charge density (a) and maximum energy (b) of the surface electrons emitted by BGs at resonance. $13^\circ$ is the optimal blaze angle for the laser wavelength. In (a), the value of charge density obtained with the sinusoidal G$30$ is added for comparison.}
\label{fig5-blazed} 
\end{figure}
From the efficiency curves reported by Edmund Optics for the BGs tested in the experiments, only the BG$13$ is expected to have a high efficiency at the laser wavelength, since its blaze wavelength is close to $800$ nm.
Fig.~\ref{fig5-blazed} shows the charge density and maximum energy measured from all the BGs irradiated at resonance ($30^\circ$), confirming that the SP excitation is indeed optimized in correspondence of the BG$13$. In particular, the charge density is $\sim 5$ times higher than what is obtained with the G$30$, because $\sim 660$ pC are typically measured in the electron bunch. On the other side, the angular divergence along both $\theta$ and $\phi$ increases (compare Table \ref{tab2-BG13-G30Alu}). The deterioration of the spatial distribution of the electron bunch is believed to depend of the Aluminum coating of the BGs. This layer could in fact suffer from early ionization by the residual laser pedestal \cite{Mordovanakis2009}, slightly altering the depth and profile of the grating during the interaction.

\begin{table}[t]
\begin{tabularx}{\columnwidth}{YYYYY}
						& G$30$ 						& G$30_\text{Alu}$ 			& BG$13$ 			& BG$13$ rev \\[1ex]
\midrule
Al thickness				& none						&$\sim 100$ nm			& $1$ $\mu$m 			& $1$ $\mu$m 	\\
profile					& \includegraphics[width=0.15\columnwidth]{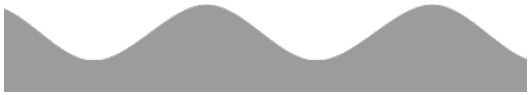}&  \includegraphics[width=0.15\columnwidth]{./grat-sin.pdf}			& \includegraphics[width=0.15\columnwidth]{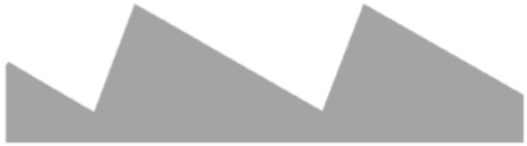}				& \includegraphics[width=0.15\columnwidth]{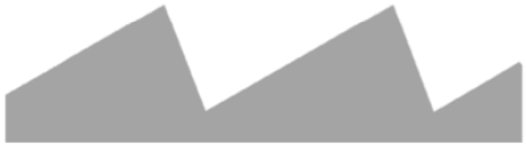}	\\[1ex]
$\theta_\text{FWHM} (^\circ)$ 	& $6.0 \pm 0.5$ 				& $7.0 \pm 0.3$			&	$9.3 \pm 0.9$		& $11.6 \pm 0.2$ \\
$\phi_\text{FWHM} (^\circ)$	& $6.5 \pm 0.5$ 				& $6.0 \pm 0.2$			&	$5.4 \pm 0.5$		& $5.8\pm 0.1$ \\
charge (pC)				& $95 \pm 5$ 			& $28 \pm 3$				&	$\mathbf{660} \pm 80$	& $19 \pm 1$ 	\\
$E_\text{peak}$			& $6 \pm 2$ 					& $4.2 \pm 0.1$			&	$7.7 \pm 0.8$		& $3.8 \pm 0.2$ \\
$E_\text{max}$				& $17 \pm 3$ 			& $10.5 \pm 0.3$			&	$\mathbf{18} \pm 2$		& $7.6 \pm 0.3$ \\
\end{tabularx}
\caption{Properties of the electron emission obtained at resonance ($30^\circ$) with the bare G$30$ and with the Aluminized gratings (G$30_\text{Alu}$, BG$13$ and BG$13$ reversed). The orientation for the BGs follows the sign convention as in Fig.~\ref{fig1-scheme}a. The results indicate that the Al coating spoils the electron acceleration, yet the optimal blaze profile in the right orientation accounts for a high amount of charge.}
\label{tab2-BG13-G30Alu}\looseness=-1 
\end{table}
Further evidence of this effect comes from the measurements on the Aluminized sinusoidal grating (G$30_\text{Alu}$), reported in Table \ref{tab2-BG13-G30Alu}. In particular, electrons are less numerous and slightly more dispersed with respect to the bare G$30$ described in the previous section. Also the maximum energy is reduced, in contrast with the theoretical model that predicts a growth of the maximum energy following the density increase ($W\propto \sqrt{n_e}$). The experimental results clearly suggest that despite providing a higher electron density, the Aluminum hampers the electron acceleration along the surface.  However,
this drawback might be mitigated on the BG$13$ because of both the deeper profile and the presence of the blaze, although the maximum electron energy remains of the same order of what is measured with the G$30$ and G$45$. 

The last column in Table \ref{tab2-BG13-G30Alu} contains the bunch properties observed when the orientation of the BG$13$ was reversed (\textit{i.e..}~the blaze angle points to the same side of the incident laser beam). Since the sawtooth profile is asymmetric, this change does not displace the diffraction orders (and indeed a bunch is still emitted along the grating surface), but it affects the grating efficiency\cite{Richardson}, as clearly demonstrated by the poorer characteristics of the electrons observed in this case. 

All these results indicate that the amount of charge and the final electron energy are strongly sensitive to the details of the grating structure, as blazed profiles achieve better results than sinusoidal gratings. Hence, engineering the target surface on a sub-micrometric scale can be exploited to optimize both the interaction and the secondary emissions also in the high intensity regime, even when the target is eventually heated to very high temperatures.

\section{Numerical results} \label{sec-simulations}
We tested and validated our experimental results with two-dimensional particle-in-cell (PIC) simulations performed with the open source code PICCANTE \cite{Sgattoni2015a} on the HPC cluster CNAF (Bologna, Italy). 
The 2D geometry is adequate to assess the main features of a surface mechanism such as the SP excitation, since it includes all the fundamental elements of the interaction. Nevertheless, 3D simulations have shown to better reproduce the fine structure of the energetic spectra and to reveal a possible correlation between the electron energy and the emission angle \cite{Fedeli2016}. 

\begin{figure}[tb]
\includegraphics[width=1.0\columnwidth]{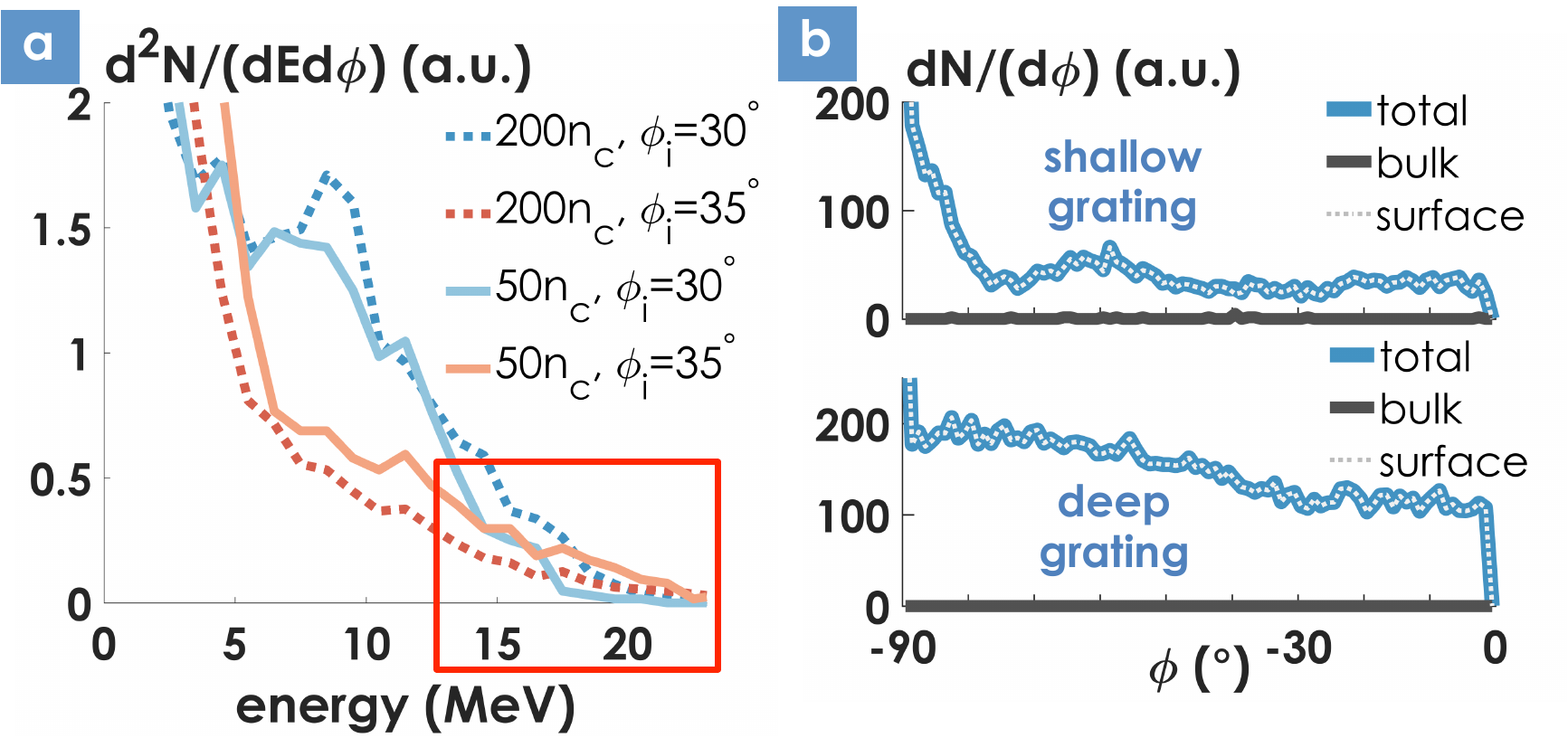}
\caption{Evaluation of the initial parameters of the 2D PIC simulations. (a) Energetic spectra of the electron emitted along the tangent of a G$30$ irradiated at $30^\circ$ (blue lines) or $35^\circ$ (red lines) of incidence, for an initial target density of $50n_c$ (solid lines) or $200n_c$ (dashed lines). Electrons above $17$ MeV (within the red box) are suppressed at $35^\circ$ in case of high density. (b) Spatial distribution of electrons (with energy above 1 MeV) emitted from the surface and the bulk  of a $5\lambda$-thick G$30$ irradiated at resonance. The grooves are $0.25\lambda$ (top) and $2\lambda$ (bottom) deep; the surface layer includes the whole groove and a further $0.5\lambda$ of thickness.  Visibly, no electrons come from the substrate.}
\label{fig6-pic-parameters} 
\end{figure}

The relevant parameters of the 2D simulations are reported in Table \ref{tab3-pic}. 
\begin{table}[bp]
\begin{center}
\begin{tabular}[b]{lcl}
Parameter		&&	values 				\\
\midrule
box size $(x,y)$						&& $100\lambda \times 100 \lambda$		\\
spatial resolution $\Delta x$, $\Delta y$ 	&& $\lambda/70$, $\lambda/40$ \\
boundaries						&& periodic 				\\[1ex]
target density $n_0$		&& $50n_c$ \\
particles per cell		&& $128$ electrons, $25$ ions \\ 
target shape			&& flat, sinusoidal/blazed grating \\
target location ($x,y$ limits)			&& $[0,$thickness$] \times [-50\lambda,50\lambda]$ \\ 
grating depth $d$		&& scan from $0.25\lambda$ to $2\lambda$	\\		
target thickness		&& scan from $1\lambda$ to $5\lambda$ \\[1ex]
laser polarization		&& 	P \\
laser $a_0$			&& 	$5$ \\
laser duration $\tau$ FWHM		&& $12 \lambda/c$\\
laser waist $w_0$		&& $5\lambda$ \\
\end{tabular}
\caption{Setup of the 2D PIC simulations.}
\label{tab3-pic}
\end{center}
\end{table}
The overdense target is placed at the center of the simulation box, with the irradiated surface at $x=0$, and extending in the $y$ direction over the whole range $-50\lambda \times 50\lambda$.
To relieve the computational load, both the density and thickness of the target are smaller than their experimental values. However, few simulations were performed with $n_0=200n_c$ and thicker targets (up to $5\lambda$). As a result of the increased density, the dispersion relation of the SP (in Eq.~\ref{res-cond}) weakly depends on the plasma density, allowing to narrow the range of incidence angles for which the electron acceleration takes place. In particular, it was possible to ascribe some energetic electrons, observed in the low density simulations for non-resonant angles, to the target heating rather than to the excitation of a SP, as shown in Fig.~\ref{fig6-pic-parameters}a. Simulations with thick targets, instead, demonstrated that the electrons accelerated in the bunch come only from the surface layer of the target, while the electrons from the bulk do not contribute to the spatial emission over the entire $\phi$ range (Fig.~\ref{fig6-pic-parameters}b). 

At the end of the simulations ($t=55\lambda/c$), the electron phase-space was analyzed to infer the energetic spectrum in the tangent direction (centered at $\phi=\arctan{(y/x)}=-88^\circ \pm 1^\circ$) and the angular distribution ($-90^\circ < \phi <0^\circ$, with the same sign convention as in Fig.~\ref{fig1-scheme}a). Only the electrons emitted in front of the target were considered ($-50<x<d/2$, where $d$ is the groove depth). Target absorption was estimated by comparing the fraction of energy possessed by all the particles in the box to the initial energy of the laser pulse.

\subsection{Role of the grating depth} \label{sec-simulations-1}
Due to manufacturing constraints, the thin gratings explored in section \ref{sec-experiments-1} had different groove depths. 
The resonance condition in Eq.~\ref{res-cond} is valid under the assumption that the groove depth is smaller than the grating period, $d \ll \Lambda$. But even within this limit, the groove depth could have a significant influence on the coupling between the target and the laser pulse, as corrugated and micro-structured targets are generally known to increase the laser target absorption because of local field enhancement \cite{Klimo2011, Andreev2011, Jiang2014}. Therefore, we tested the role of the grating depth by running several simulations, where firstly we varied the groove depth while keeping the grating period constant.

\begin{figure}[htb]
\includegraphics[width=1.0\columnwidth]{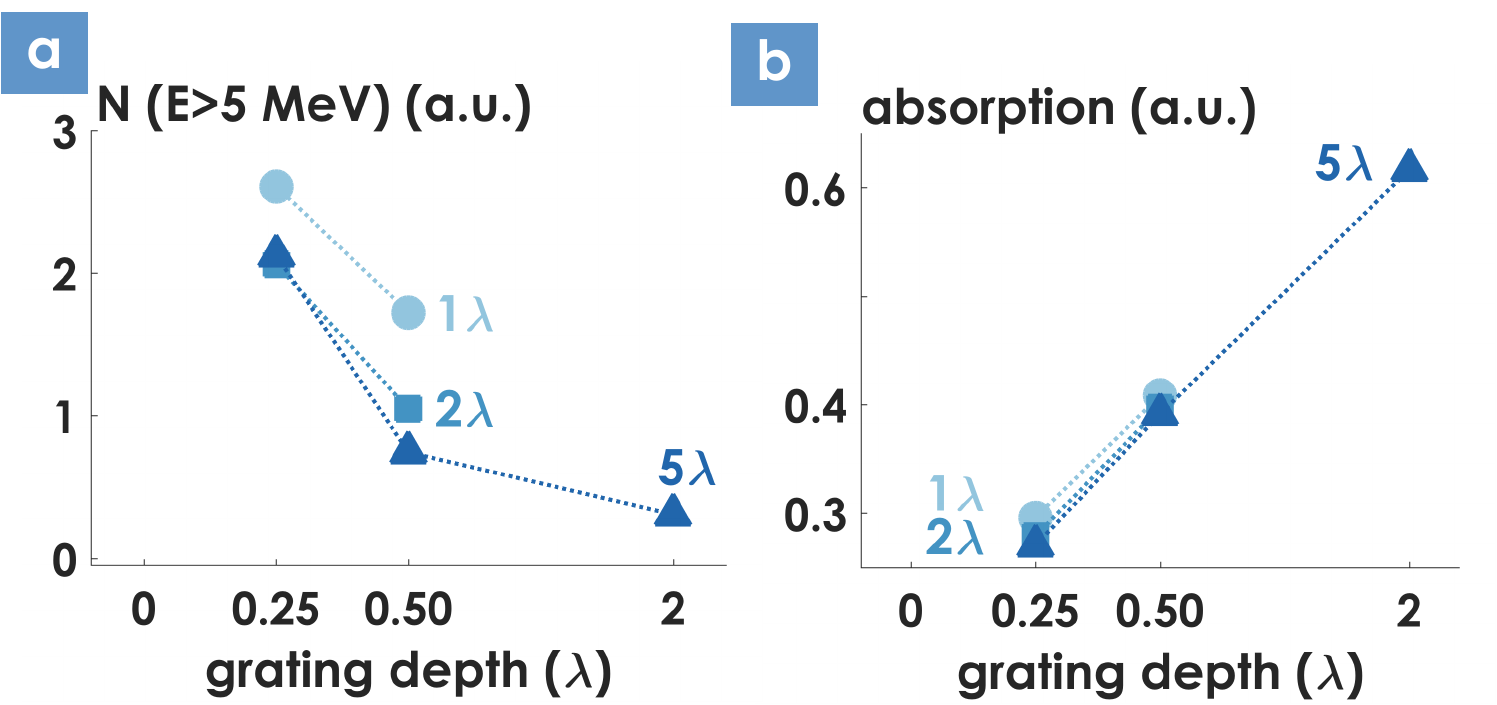}
\caption{Effect of increasing the groove depth in a G$30$ irradiated at $30^\circ$: (a), number of electrons beyond $5$ MeV accelerated at tangent and (b), grating absorption. Markers, colors and labels specify the target thickness. Deep grooves reduce the efficiency of the SP excitation, despite increasing the target absorption. Recirculation effects might account for the larger electron number observed with thinner gratings.}
\label{fig7-pic-depth-scan} 
\end{figure}
Fig.~\ref{fig7-pic-depth-scan} illustrates how increasing the groove depth spoils the surface electron acceleration, despite increasing the target absorption. In these simulations, the target was a G$30$ (\textit{i.e.}~$\Lambda=2\lambda$) irradiated at resonance; the target thickness is indicated next to the curves. Fig.~\ref{fig7-pic-depth-scan}a shows that the number of electrons with energy above $5$ MeV emitted along the grating surface drastically decreases when the groove depth is beyond $0.5\lambda$, whereas the absorption (Fig.~\ref{fig7-pic-depth-scan}b) increases. Both results show that although deep grooves cause the shadow effect to hamper the SP excitation\cite{Raynaud2007}, geometrical effects are still able to increase the absorption\cite{Ceccotti2013, Bigongiari2013, Bigongiari2012}. 

\begin{figure}[tb]
\includegraphics[width=1.0\columnwidth]{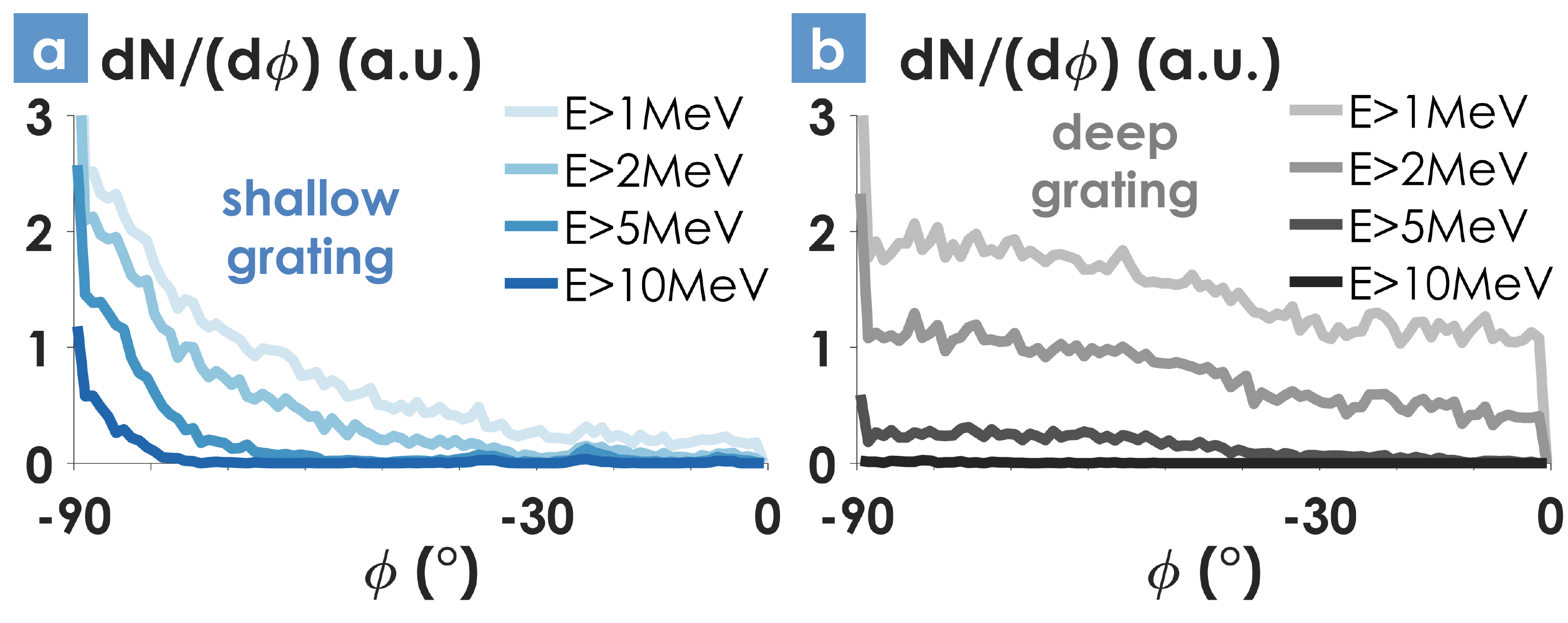}
\caption{Spatial distribution of the electron emission, for a G$30$ irradiated at resonance with different depth-thickness configurations ($0.25\lambda-1\lambda$ in (a) and $2\lambda-5\lambda$ in (b)). The legends report the energy filters applied to the electron population. Only the shallow grating exhibits energetic electron along the surface ($\phi < -80^\circ$).}
\label{fig8-pic-philineout} 
\end{figure}

However, it is worth noticing that the high absorption is largely due to low-energy electrons, as Fig.~\ref{fig8-pic-philineout} shows. In fact, the full spatial distribution of the electrons emitted from the deep G$30$ (Fig.~\ref{fig8-pic-philineout}b) indicates that most electrons are below $5$ MeV of energy; the shallow grating, on the contrary, exhibits a distinctive emission in the tangent direction ($\phi \simeq -90^\circ)$ of electrons between $5$ and $10$ MeV (Fig.~\ref{fig8-pic-philineout}a). This allows to exclude that a SP is excited with deep grooves, as confirmed also by analyzing the electron distribution obtained with different incidence angles (not shown here, for brevity). 

Since the shadow effect depends on both the grating depth and period, the groove depth where the SP excitation becomes inefficient is expected to be different for each resonant angle investigated in section \ref{sec-experiments-1}. 
\begin{figure}[t]
\includegraphics[width=1.0\columnwidth]{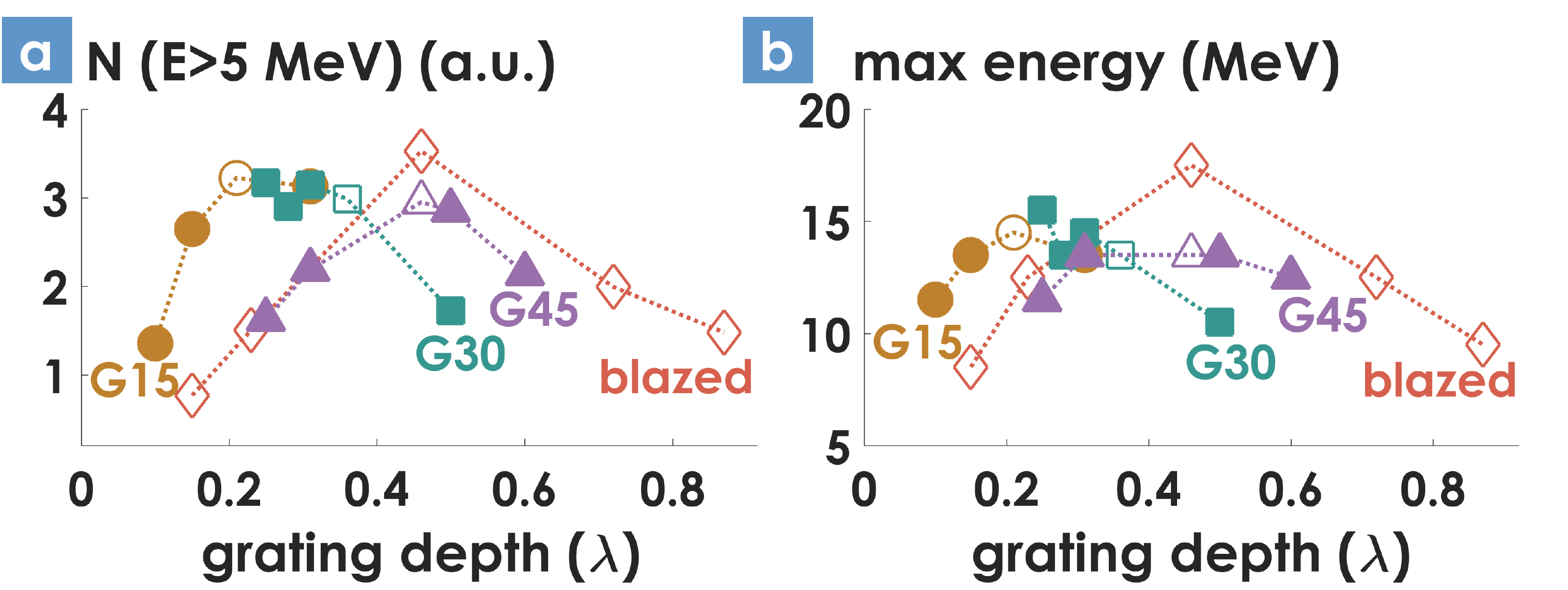}
\caption{Depth scan on gratings with different periods: both the electrons at tangent beyond $5$ MeV (a) and the maximum energy attained by the spectra (b) are optimized within a certain range of depth. Empty points represents the simulations where the groove depth corresponded to the experimental value ($0.21\lambda$ for the G$15$, $0.36\lambda$ for the G$30$, $0.46\lambda$ for the G$45$ and the values fixed by the blaze angles for the BGs). The bulk thickness was $2\lambda$ in all these simulations.}
\label{fig9-pic-depth-comparison} 
\end{figure}
Indeed, Fig.~\ref{fig9-pic-depth-comparison} confirms that each grating requires a specific groove depth to optimize the charge and energy of the SP-accelerated electrons. In these simulations, all gratings were irradiated at resonance, and the groove depth was varied for both the sinusoidal gratings (G$15$, G$30$ and G$45$) and the blazed gratings (in this case the blaze angle automatically determines the groove depth). Both graphs show that increasing the resonant angle (\textit{i.e.}~the grating period) requires to increase the groove depth; however, the optimal performances of all sinusoidal gratings are comparable, as it was found in the experiment (compare the charge in Fig.~\ref{fig3-spatial-properties}b or the maximum energy in Fig.~\ref{fig4-spectra}c). 
Moreover, the simulations where we modeled the depth of the sinusoidal gratings with the values as in the experiments confirm that there are no significant differences among the surface electron emission obtained with these targets (neither in the energetic spectra nor in the spatial distributions, which are not shown here for brevity). 
\looseness=-1 Further improvement of the grating efficiency is once again obtained with the suitable blazed grating, the BG$13$.

In a final set of simulations, we investigated the role of the number of grating periods illuminated by the laser pulse, which depends on both the groove spacing and the incidence angle. Therefore, we tested a G$45$ irradiated at $45^\circ$ of incidence, by a laser pulse whose focal spot was adapted to cover the same number of periods as the G$30$ irradiated at $30^\circ$ (by a beam with waist $5\lambda$). The peak intensity was kept at $a_0=5$. As result, the two configurations produced equivalent electron emissions at the grating surface, provided that the depth of the G$45$ was larger with respect to the G$30$. This study pointed out that the number of grating periods is not a crucial parameter required to optimize the electron acceleration.

\subsection{Scan of the laser conditions} \label{sec-simulations-2}
With another set of numerical simulations, we covered the SP-driven electron acceleration with different laser parameters. The goal was both to evaluate the possible influence of some interaction conditions, and to explore the scaling laws of the acceleration mechanism.

\begin{figure}[t]
\includegraphics[width=0.85\columnwidth]{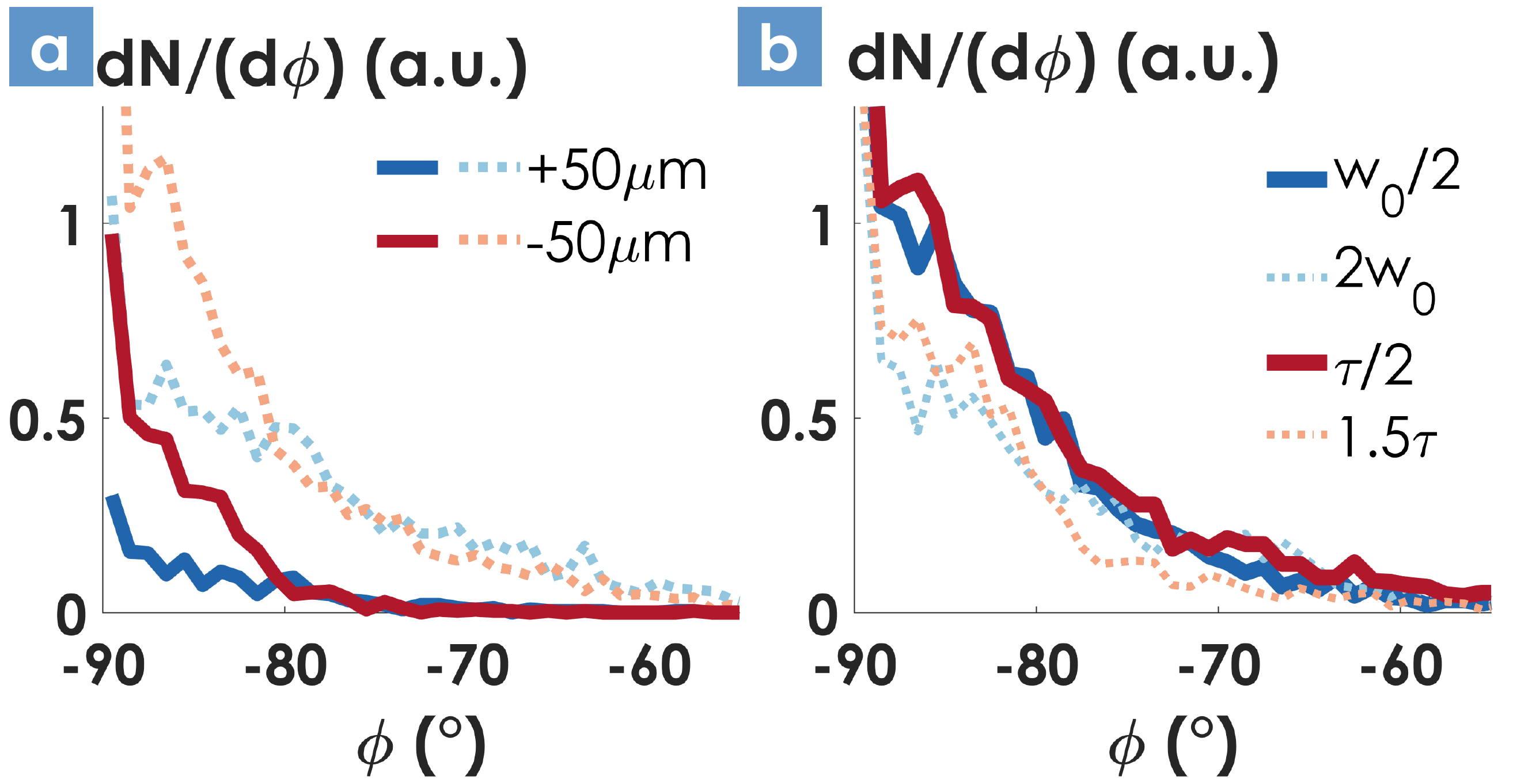}
\caption{Scan of the laser parameters on a G$30$ irradiated at resonance. The electron spatial distributions are shown for: (a), different grating positions along the focal axis, for electrons above $5$ (dashed lines) or $10$ MeV (solid lines); (b), different beam waists $w_0$ or pulse durations $\tau$ at fixed laser energy. In (b), the configurations at higher intensity are represented with solid lines.}
\label{fig10-pic-laser-scan} 
\end{figure}

In the first case we performed a scan of the grating position along the axis of the laser beam, and of the grating phase with respect to the focal point (whether the center of the focal spot hits the sinusoidal profile on a peak, on a valley, or somewhere between them). The target implemented in all these simulations was a G$30$ irradiated at resonance, with a groove depth $0.36\lambda$ and a substrate thickness of $2\lambda$. As result, we found that the grating phase has no influence on the electron acceleration over the entire $\phi$ range. Fig.~\ref{fig10-pic-laser-scan}a, instead, shows that a $\sim \pm 50$ $\mu$m shift of the focal position leads to $\sim 40$\% fluctuations on the charge emitted along the tangent and, from the spectra, $\pm 2$ MeV on the maximum energy. 

We also varied either the beam waist or the pulse duration, keeping the laser energy constant. The electron spatial distribution, illustrated in Fig.~\ref{fig10-pic-laser-scan}b, indicates that the configurations at higher intensity (small focal spot or short pulse, solid lines) are the most favorable for the electron acceleration (note that fewer irradiated lines correspond here to a higher laser intensity, whereas in the simulation described in the previous section the peak intensity was fixed).

\begin{figure}[t]
\includegraphics[width=1.0\columnwidth]{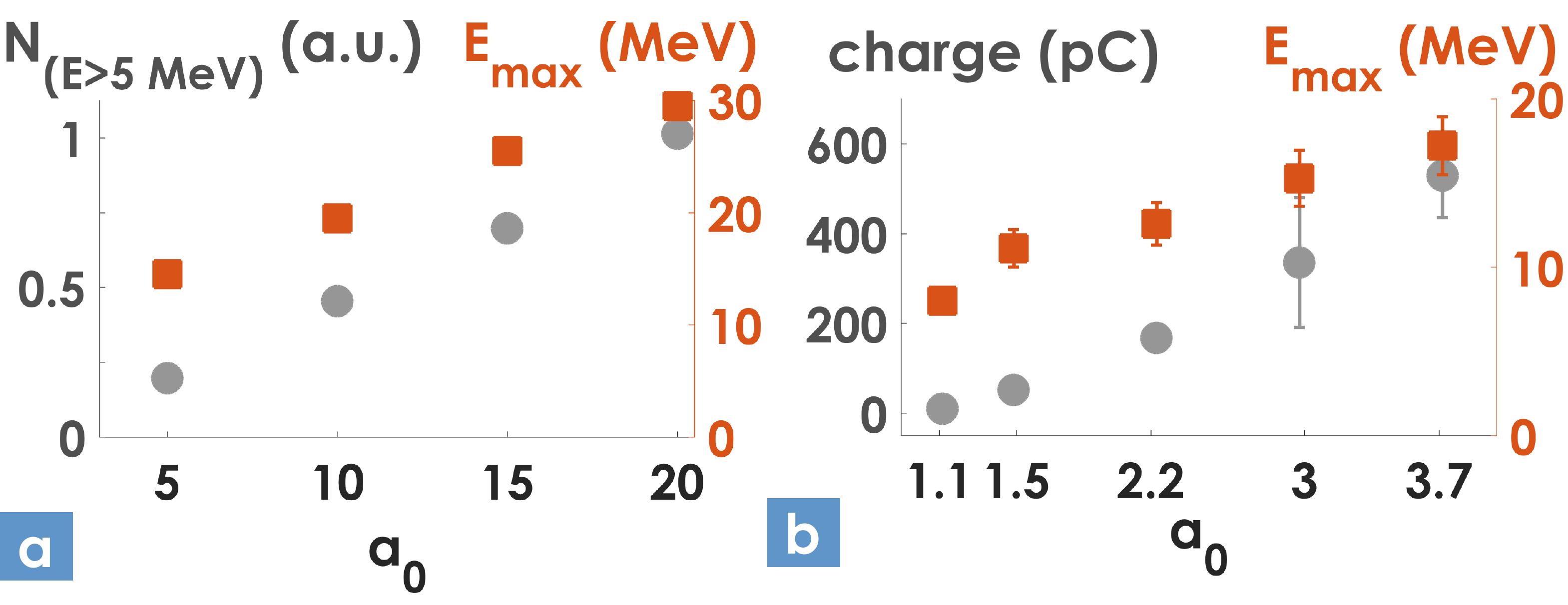}
\caption{(a) Number of electrons above $5$ MeV emitted at tangent and maximum energy as a function of the laser peak intensity, from PIC simulations. (b) Charge amount and $E_\text{max}$ measured on the electron bunch emitted from a BG$13$ at resonance, when varying the energy of the laser pulse on the UHI-100 facility.}
\label{fig11-laser-a0} 
\end{figure}
As a final consideration, the theoretical model\cite{Riconda2015} briefly presented in section \ref{sec-theory} predicts for the electron energy a scaling proportional to $a_0$ ($W \propto a_{SP} \sim a_0$), but makes no predictions with respect to the amount of charge emitted along the grating surface. 
Fig.~\ref{fig11-laser-a0}a illustrates that also this quantity exhibits a linear trend with increasing laser intensities. The results of the numerical simulations, in particular, are compared to the experimental measurements performed on UHI-100 (Fig.~\ref{fig11-laser-a0}b), for $a_0$ between $1.1$ and $3.7$. The target is a BG$13$, irradiated at resonance. Despite the necessary caution in comparing the signal emitted by the Lanex screens and the result of 2D PIC simulations, the linear scaling is also supported by the experimental points. Further accuracy would surely result from evaluating the dependence of the simulations on the target density and the 3D geometry. 
If confirmed, these trends would suggest that the total energy of the surface electrons scales linearly with the laser energy, implying a quasi-constant efficiency of the acceleration mechanism.

\section{Conclusions}
This article provides numerous evidence of SPs excitation in the relativistic regime, by analyzing its role in the acceleration of intense electron bunches along the target surface.

Gratings irradiated at the resonant angle for SP excitation predicted by the linear theory exhibit a bright and highly-directional emission of energetic electrons, whose properties dramatically worsen when changing the incidence angle or spoiling the temporal contrast of the laser pulse. Flat foils rather produce a $20$-time weaker electron cloud around the specular reflection of the laser beam.
Varying the grating period, profile and material has allowed us to demonstrate the robustness of the acceleration mechanism and to identify useful guidelines for its optimization. In particular, the most suitable blazed grating produce $\sim 700$ pC of charge with an energetic spectrum centered at $\sim 10$ MeV and reaching $18$ MeV of cutoff. 
The experiments also indicate that dielectric materials give place to electron beams with higher charge and lower divergence with respect to metallic gratings. 

\looseness=-1 With 2D PIC simulations we have explored the role of various parameters of the laser-grating interaction on both the surface electron acceleration and the target absorption, such as the number and position of grating periods irradiated by the laser pulse, the characteristics of the laser pulse, and the grating depth. In particular, there exists an optimal groove depth, which depends on the resonant angle, where the electron acceleration is most efficient; this agrees with the SP theory which requires shallow gratings to derive the resonance condition. Inversely, complex geometrical effects account for the high absorption achieved with deep grooves, although the electron emission is neither collimated nor energetic in this case. 

The observation of SP-accelerated electron bunches demonstrates the feasibility of SP excitation in the relativistic regime, warming up both the theoretical and experimental investigation of high field Plasmonics \cite{Fedeli2016b}. 
Moreover, the accurate characterization the electron emission is the first step towards their promising application as a bright, laser-synchronized, ultra-short electron source at modest energies, potentially suitable for high-repetition rate schemes\cite{Prencipe2017}. Indeed, although the energetic spectra are far from being mono-chromatic, the peak energies belong to a range hardly attainable with the laser wakefield mechanism, the charge amounts are much higher, and the simple interaction geometry supports the integration of these electron sources in more complex target structures.

\begin{acknowledgments}
We acknowledge M. Kv\v{e}to\v{n} (HoloPlus, Prague, Czech Republic) for the prompt and accurate manufacturing of the thin gratings; Jean-Philippe Larbre (ELYSE, Universit{\'e}  Paris Sud, Orsay, France) for his contribution during the calibration of the Lanex screen; the computational support from HPC Cluster CNAF (Bologna, Italy), with particular thanks to S. Sinigardi for his precious assistance; F. Amiranoff (LULI, UPMC, Paris, France) for the invaluable discussions on the theory of SP excitation and electron acceleration. G.C.  acknowledges financial support from the Universit\'e Franco-Italienne (Vinci program, grant No. C2-92).
\end{acknowledgments}

\bigskip \bigskip

%

\end{document}